\documentclass[aps,prd,eqsecnum,preprint,tightenlines,nofootinbib,showpacs,floatfix]{revtex4-1}
\usepackage{graphicx,latexsym}

\def\bea{\begin{eqnarray}}
\def\eea{\end{eqnarray}}
\def\be{\begin{equation}}
\def\ee{\end{equation}}

\newcommand{\Pminus}{{\cal P}^-}
\newcommand{\Nmax}{N_{\rm max}}

\begin{document}

\title{Light-front $\phi_2^4$ theory with sector-dependent mass}
\author{Sophia S. Chabysheva}
\author{John R. Hiller}
\affiliation{Department of Physics and Astronomy\\
University of Minnesota-Duluth \\
Duluth, Minnesota 55812}

\date{\today}

\begin{abstract}
As an extension of recent work on two-dimensional light-front $\phi^4$ theory,
we implement Fock-sector dependence for the bare mass.  Such 
dependence should have important consequences for the convergence of
nonperturbative calculations with respect to the level of Fock-space
truncation.  The truncation forces the self-energy corrections to be
sector-dependent; in particular, the highest sector has no self-energy
correction.  Thus, the bare mass can be considered sector dependent
as well.  We find that, although higher Fock sectors have a larger
probability, the mass of the lightest state and the value of the
critical coupling are not significantly affected.  This implies that
coherent states or the light-front coupled-cluster method may be
required to properly represent critical behavior.

\end{abstract}

%

\maketitle

\section{Introduction}
\label{sec:Introduction}

In a recent calculation~\cite{phi4sympolys}, two-dimensional $\phi^4$ theory was
solved for the lowest mass eigenstates of the light-front Hamiltonian.  The
eigenstates were represented by truncated Fock-state expansions, with momentum-space
wave functions as the coefficients.  This work included estimation of
the critical coupling, where the $\phi\rightarrow -\phi$ symmetry of the theory
is broken~\cite{Chang}.  At this critical coupling, one would expect that the 
probabilities for the higher Fock sectors, computed as integrals of the 
squares of the Fock wave functions, would increase dramatically.  In particular,
the probability for the one-particle sector of the lowest massive state
should go to zero.  This was not observed.

The expectation that the one-particle probability would go to zero is
important for the calculation of the connection between 
equal-time~\cite{RychkovVitale,LeeSalwen,Sugihara,SchaichLoinaz,Bosetti,Milsted}
and light-front estimates~\cite{phi4sympolys,VaryHari} of the critical coupling.  
This is determined by the relationship between the different mass renormalizations
in the two quantizations~\cite{SineGordon}, which is fixed by
tadpole contributions computed from the vacuum expectation value
of $\phi^2$.  The behavior of this vacuum expectation value is
dominated by the product of the one-particle probability times
the logarithm of the mass~\cite{phi4sympolys}.  The mass goes to zero at the critical
coupling, making zero probability a necessity for a finite result.

In the previous work, the explanation proposed for this apparent
paradox was that the calculation did not use sector-dependent
bare masses.  This meant that the highest Fock sector kept in
the calculation used a fixed bare mass even as the eigenstate
mass approached zero.  Excitation of such Fock states is then
very unlikely.

The use of sector-dependent bare parameters, or `sector-dependent
renormalization' as it is usually called, has a long 
history~\cite{SecDep-Wilson,HillerBrodsky,Karmanov}.  
A Fock-space truncation forces self-energy corrections and
vertex corrections to be different in different Fock sectors.
This makes sector-dependent counterterms a natural choice.
In addition, the truncation causes divergences that would
have been canceled by contributions from higher Fock states
that are now absent.  Sector-dependent counterterms can
take these divergences into account.  However, in at 
least some theories, the sector-dependent bare couplings
can lead to inconsistencies in the interpretation of 
wave functions and Fock-sector probabilities~\cite{SecDep}.
Thus, use of sector-dependent bare masses can be a 
compromise.  Of course, for two-dimensional $\phi^4$ theory,
divergences are not the issue, and it is only near the 
critical coupling where sector-dependent masses could
be a useful approximation, as already indicated by
some preliminary work~\cite{LFCCphi4}.

The use of light-front quantization~\cite{LFreviews} is 
important for its simple vacuum and well-defined 
Fock state expansions as well as for the separation
of relative and external momenta.  We define light-front
coordinates~\cite{Dirac} as $x^\pm=t\pm z$, with
$x^+$ the light-front time.  The light-front energy
is then $p^-=E-p_z$, and the light-front momentum is
$p^+=E+p_z$.  In what follows, we will drop the superscript
from $p^+$ to simplify the notation.  The inner product
between momentum and position is $p\cdot x=\frac12(p^-x^++px^-)$,
and the mass-shell condition is $p^-=m^2/p$.  The
light-front Hamiltonian eigenvalue problem is then
$\Pminus|\psi(P)\rangle=\frac{M^2}{P}|\psi(P)\rangle$,
with $|\psi(P)\rangle$ the eigenstate with mass $M$ and
light-front momentum $P$.  The eigenstate is expanded
in a set of Fock states, which converts the formal
eigenvalue problem into a system of equations for
the Fock-state wave functions.  Numerical approximations
then transform this system into a matrix eigenvalue
problem.  Our chosen numerical approximation is an
expansion of the wave functions in terms of symmetric
multivariate polynomials~\cite{GenSymPolys}.

The content of the remainder of the paper is as follows.
Section~\ref{sec:phi4theory} provides a brief introduction
to $\phi^4$ theory and formulates the system of
equations for the Fock-state wave functions.  These
equations are modified in Sec.~\ref{sec:secdepmass} to
accommodate a sector-dependent bare mass; the results
from their solution are presented and discussed.
The work is summarized briefly in Sec.~\ref{sec:summary}.  
Details of the numerical methods are left to an Appendix.

\section{Light-front $\phi^4$ theory}
\label{sec:phi4theory}

The Lagrangian for $\phi^4$ theory is
\be
{\cal L}=\frac12(\partial_\mu\phi)^2-\frac12\mu^2\phi^2-\frac{\lambda}{4!}\phi^4,
\ee
with $\mu$ the bare mass.  The light-front Hamiltonian density is
\be
{\cal H}=\frac12 \mu^2 \phi^2+\frac{\lambda}{4!}\phi^4.
\ee
The field $\phi$ is expanded in terms of creation and annihilation
operators $a^\dagger(p)$ and $a(p)$ as
\be \label{eq:mode}
\phi(x^+=0,x^-)=\int \frac{dp}{\sqrt{4\pi p}}
   \left\{ a(p)e^{-ipx^-/2} + a^\dagger(p)e^{ipx^-/2}\right\}.
\ee
The operators obey the commutation relation
\be
[a(p),a^\dagger(p')]=\delta(p-p').
\ee
Substitution of the mode expansion and
integration of the Hamiltonian density with respect to $x^-$ yields
the light-front Hamiltonian $\Pminus=\Pminus_{11}+\Pminus_{22}+\Pminus_{13}+\Pminus_{31}$,
with
\bea \label{eq:Pminus11}
\Pminus_{11}&=&\int dp \frac{\mu^2}{p} a^\dagger(p)a(p),  \\
\label{eq:Pminus22}
\Pminus_{22}&=&\frac{\lambda}{4}\int\frac{dp_1 dp_2}{4\pi\sqrt{p_1p_2}}
       \int\frac{dp'_1 dp'_2}{\sqrt{p'_1 p'_2}} 
       \delta(p_1 + p_2-p'_1-p'_2) a^\dagger(p_1) a^\dagger(p_2) a(p'_1) a(p'_2), \\
\label{eq:Pminus13}
\Pminus_{13}&=&\frac{\lambda}{6}\int \frac{dp_1dp_2dp_3}
                              {4\pi \sqrt{p_1p_2p_3(p_1+p_2+p_3)}} 
     a^\dagger(p_1+p_2+p_3)a(p_1)a(p_2)a(p_3), \\
\label{eq:Pminus31}
\Pminus_{31}&=&\frac{\lambda}{6}\int \frac{dp_1dp_2dp_3}
                              {4\pi \sqrt{p_1p_2p_3(p_1+p_2+p_3)}} 
      a^\dagger(p_1)a^\dagger(p_2)a^\dagger(p_3)a(p_1+p_2+p_3).
\eea

The eigenstate of $\Pminus$, with eigenvalue $M^2/P$,
can be expressed as an expansion 
\be \label{eq:FSexpansion}
|\psi(P)\rangle=\sum_m P^{\frac{m-1}{2}}\int\prod_i^m dy_i 
       \delta(1-\sum_i^m y_i)\psi_m(y_i)|y_i;P,m\rangle
\ee
in terms of Fock states
\be
|y_i;P,m\rangle=\frac{1}{\sqrt{m!}}\prod_{i=1}^m a^\dagger(y_iP)|0\rangle,
\ee
where the coefficient $\psi_m$ is the wave function for the
Fock sector with $m$ constituents.  The wave function depends
on the momentum fractions $y_i\equiv p_i/P$, which are
boost invariant, unlike the individual momenta $p_i$.
The leading factor of $P^{\frac{m-1}{2}}$ allows the
normalization of the wave functions to be independent of $P$;
we require $\langle\psi(P')|\psi(P)\rangle=\delta(P-P')$,
which yields
\be \label{eq:normalization}
1=\sum_m \int\prod_i^m dy_i \delta(1-\sum_i^m y_i)|\psi_m(y_i)|^2.
\ee
The probability of the $m$th Fock sector is then just
$\int\prod_i^m dy_i \delta(1-\sum_i^m y_i)|\psi_m(y_i)|^2$.
Because the Hamiltonian changes particle number by zero or
two, never an odd number, the sum over Fock sectors is either
even or odd, depending on which state is chosen as the
lowest Fock state.\footnote{This is, of course, a consequence of
the fundamental $\phi\rightarrow-\phi$ symmetry of the
original Lagrangian.}  We will focus on the odd case.

The eigenvalue problem becomes a coupled system of
equations for the wave functions:
\bea \label{eq:coupledsystem}
\lefteqn{\sum_i^m \frac{\mu^2}{y_i }\psi_m(y_i)
+\frac{\lambda}{4\pi}\frac{m(m-1)}{4\sqrt{y_1y_2}}
        \int\frac{dx_1 dx_2 }{\sqrt{x_1 x_2}}\delta(y_1+y_2-x_1-x_2) \psi_m(x_1,x_2,y_3,\ldots,y_m)}&& 
        \nonumber \\
&& +\frac{\lambda}{4\pi}\frac{m}{6}\sqrt{(m+2)(m+1)}\int \frac{dx_1 dx_2 dx_3}{\sqrt{y_1 x_1 x_2 x_3}}
        \delta(y_1-x_1-x_2-x_3)\psi_{m+2}(x_1,x_2,x_3,y_2,\ldots,y_m) \nonumber \\
&& +\frac{\lambda}{4\pi}\frac{m-2}{6}\frac{\sqrt{m(m-1)}}{\sqrt{y_1y_2y_3(y_1+y_2+y_3)}}
          \psi_{m-2}(y_1+y_2+y_3,y_4,\ldots,y_m)=M^2\psi_m(y_i).
\eea
This is an infinite system and requires some form of truncation
before a numerical solution can be attempted.  The standard truncation
is a Fock-space truncation to some maximum number of constituents $\Nmax$.
However, as discussed in the Introduction, this causes self-energy 
contributions to become sector-dependent.
For the sector with $m=\Nmax$, there is no self-energy because no
loop corrections are allowed; any intermediate states would have more than
$\Nmax$ constituents.  Therefore, the bare mass in the top sector is
reasonably equal to the physical mass $M$ of the lowest state.  As we step
down from the top sector, the complexity of the self-energy contributions
steadily increases, and the bare mass can be adjusted to compensate.

\section{Sector-dependent mass}  
\label{sec:secdepmass}

To implement a sector-dependent bare mass, we replace $\mu$ in the first term
of (\ref{eq:coupledsystem}) by $\mu_m$ and compute the $\mu_m$ for a given 
eigenmass $M$ by steadily increasing $\Nmax$.  For $\Nmax=1$,
we have immediately that $\mu_1=M$ and $|\psi(P)\rangle=a^\dagger(P)|0\rangle$.
For $\Nmax=3$, we set $\mu_3=M$ and solve the following two equations
for $\mu_1$ and $\psi_3/\psi_1$:
\be
\mu_1^2\psi_1
+\frac{\lambda}{4\pi}\frac{1}{\sqrt{6}}\int \frac{dx_1 dx_2 dx_3}{\sqrt{x_1 x_2 x_3}}
        \delta(1-x_1-x_2-x_3)\psi_3(x_1,x_2,x_3)=M^2 \psi_1
\ee
\bea
\sum_i^3 \frac{M^2}{y_i}\psi_3(y_1,y_2,y_3)
   &+&\frac{\lambda}{4\pi}\frac{3}{2\sqrt{y_1y_2}}
        \int\frac{dx_1 dx_2 }{\sqrt{x_1 x_2}}\delta(y_1+y_2-x_1-x_2) \psi_3(x_1,x_2,y_3) 
        \nonumber \\
&& +\frac{\lambda}{4\pi}\frac{1}{\sqrt{6}}\frac{1}{\sqrt{y_1y_2y_3}}
          \psi_1=M^2\psi_3(y_1,y_2,y_3).
\eea
For $\Nmax=5$, we set $\mu_3$ to the value of $\mu_1$ obtained for $\Nmax=3$,
set $\mu_5=M$, and solve a system of three equations for $\mu_1$, $\psi_3/\psi_1$, and $\psi_5/\psi_1$.
We continue in this manner until $\mu_1$ has converged with respect to
the Fock-space truncation fixed by $\Nmax$.

\begin{figure}
\vspace{0.2in}
\centerline{\includegraphics[width=15cm]{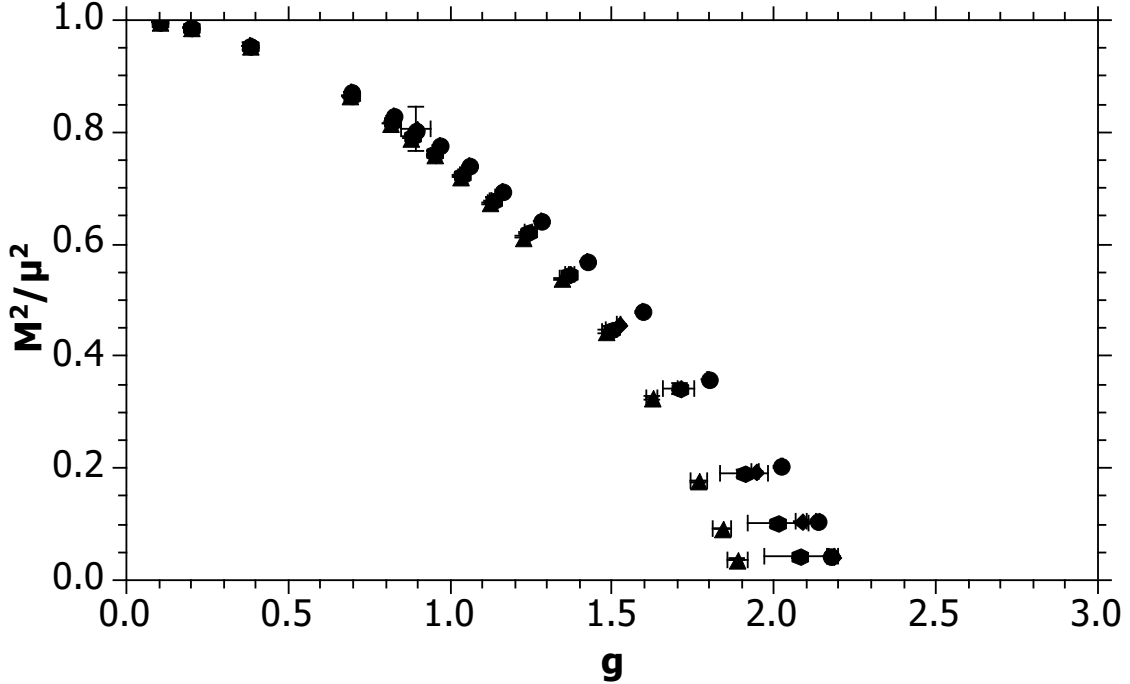}}
\caption{\label{fig:extrap}
Lowest mass eigenvalue for odd numbers of constituents
for different Fock-space truncations to three (circles),
five (triangles), seven (diamonds), and nine (hexagons)
constituents.  The values of $M^2/\mu^2$ and $g\equiv\lambda/(4\pi\mu^2)$ are
obtained as extrapolations in the orders of basis
polynomials, and the error bars estimate the range
of fits for the extrapolations.
}
\end{figure}

As described in the Appendix, the equations are solved numerically,
with the wave functions expanded in a polynomial basis~\cite{GenSymPolys}.
The principal result of the calculation is the plot of the eigenvalues 
versus coupling strength in Fig.~\ref{fig:extrap}.  These values are
extrapolated in the polynomial basis size for each Fock sector, and the
Fock-space truncation is varied from $\Nmax=3$ to 9.  With respect to
the Fock-space truncations, the results converge, in an oscillatory
fashion, to within the numerical error at a given truncation.

\begin{figure}
\vspace{0.2in}
\centerline{\includegraphics[width=15cm]{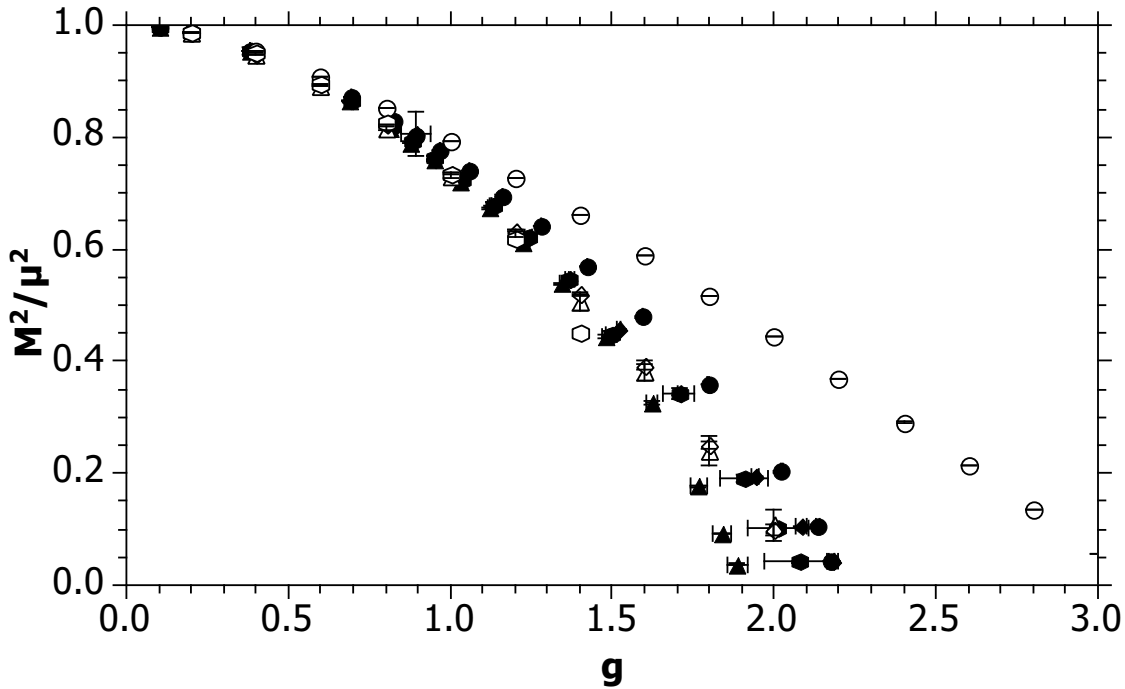}}
\caption{\label{fig:extrap-all}
Same as Fig.~\ref{fig:extrap}, but with the standard
parameterization results (open circles, triangles, and
diamonds) of \protect\cite{phi4sympolys},
which include up to seven constituents,
and the light-front coupled cluster results (open hexagons) of \protect\cite{LFCCphi4}
added for comparison.  Without sector-dependent masses, the
Fock state expansion converges more quickly, and the five and
seven-constituent results are nearly identical, between themselves
and with the nine-constituent sector-dependent results.
}
\end{figure}

These results are consistent with those from the standard parameterization,
with no sector dependence in the bare mass, as reported in \cite{phi4sympolys}.
This can be seen in Fig.~\ref{fig:extrap-all}, where the previous
results are added to the plot from Fig.~\ref{fig:extrap}.  The
critical coupling, as indicated by the point where $M^2$ reaches zero, is again
estimated to be 2.1.  The sector-dependent results do converge more slowly;
they require $\Nmax=9$ compared to the $\Nmax=5$ required for the
standard parameterization.  This is to be expected, even desired, because
we expect that higher Fock states should become more important as the
critical coupling is approached.

\begin{figure}
\vspace{0.2in}
\centerline{\includegraphics[width=15cm]{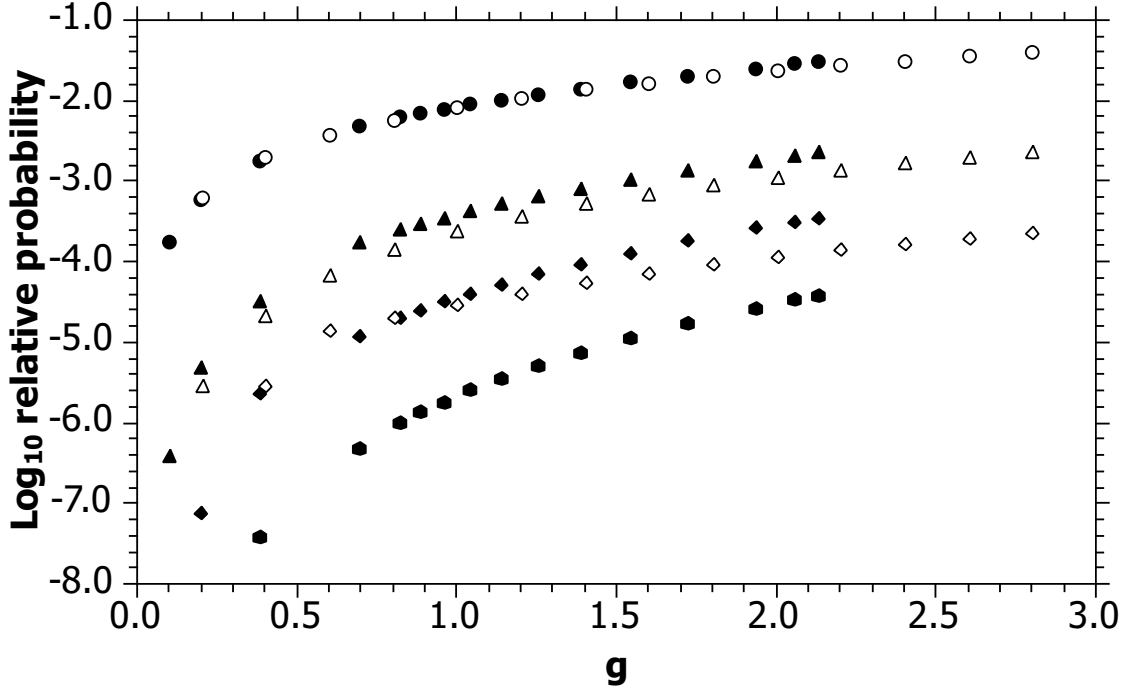}}
\caption{\label{fig:relprob} 
Relative Fock-sector probabilities for the three, five, seven
and nine-constituent sectors when the truncation is $\Nmax=9$.
}
\end{figure}

To understand what might be happening in the structure of the
eigenstate, we plot the relative Fock-sector probabilities in
Fig.~\ref{fig:relprob}.  These are computed as 
$\int\prod_i^m dy_i \delta(1-\sum_i^m y_i)|\psi_m(y_i)|^2/|\psi_1|^2$,
for both the sector-dependent and standard parameterizations.
In the former case, $\Nmax=9$ and in the latter, $\Nmax=7$.
For the sector-dependent calculations, results do not
extend beyond the critical coupling, because negative $M^2$
is ill-defined for sector-dependent renormalization; the
bare mass in the highest Fock sector would then be imaginary.
The relative probabilities are essentially the same in
the three-body Fock sector, indicating full convergence
with respect to the Fock-space truncation.  In Fock sectors
with five and seven constituents, the relative probability
for the sector-dependent case rises above the probability
in the standard case as the critical coupling is approached.
The greater probability is expected; however, the full
expectation was that these probabilities would have a
much more striking increase.  In fact, as relative probabilities,
they will tend to infinity as the one-body probability
$|\psi_1|^2$ goes to zero, and obviously this is not
happening, and the original hypothesis, that sector-dependent
bare masses would resolve the paradox, must be incorrect.

The finite one-body probability also prevents any improvement
in the calculation of the difference in mass renormalization
between equal-time and light-front quantization, as attempted
in \cite{phi4sympolys}.  Therefore, no new calculation
is attempted here.

\section{Summary}
\label{sec:summary}

Contrary to expectations, we have found that a sector-dependent
bare mass does not provide any significant improvement in the
solution of $\phi_2^4$ theory.  The sector dependence does
allow higher Fock states to have a larger contribution, but
the contribution to the lowest massive state is not large in 
an absolute sense, and the one-body contribution remains dominant, 
even at the critical coupling.  Convergence of the bare mass in 
the lowest sectors, as the Fock-space truncation is relaxed, is 
quite rapid, as can be seen in Fig.~\ref{fig:mu2vsNmax}.

There is, however, a hint as to what might be needed in
earlier work~\cite{LFCCphi4} that explored the light-front coupled-cluster
(LFCC) method~\cite{LFCC}.  In this method, all of the higher
Fock states are kept.  To keep the calculation finite in size,
the relationship between Fock wave functions is truncated, so that
wave functions of the higher Fock states are related to those
of the lower states in a simple way.  The wave functions are then
determined by a nonlinear equation that sums over contributions 
from all Fock states.  In this calculation, a relative 
probability shows a rapid increase, in Fig.~5 of \cite{LFCCphi4},
although at an unexpected value of the coupling.\footnote{The
unexpected value of $g\simeq 1.5$
may be due to the simplicity of the particular LFCC approximation;
a higher-order LFCC approximation should be investigated.}
The hint is that coherent effects across all of Fock space are 
important, something that ordinary Fock-space truncation would not
be able to reproduce.  That this would happen at the phase
transition to broken symmetry is actually not surprising.

\acknowledgments
This work was supported in part by 
the Minnesota Supercomputing Institute through
allocations of computing resources.

\appendix

\section{Numerical methods}  \label{sec:appendix}

The coupled system (\ref{eq:coupledsystem}), modified to use sector-dependent
masses, is solved numerically, with each wave function expanded in a 
basis of symmetric multivariate polynomials~\cite{GenSymPolys} $P_{ki}^{(m)}(y_1,\ldots,y_m)$.
Here $k$ is the order and $m$ the number of momentum fractions; the index $i$
differentiates between linearly independent polynomials of the same order.
The expansion for a wave function is
\be \label{eq:expansion}
\psi_m(y_1,\ldots,y_m)=\sqrt{y_1 y_2\cdots y_m}\sum_{ni} c_{ni}^{(m)} P_{ni}^{(m)}(y_1,\ldots,y_m)
\ee
Projection of the system of equations onto the basis functions then yields
a system of matrix equations
\be \label{eq:matrixequations}
\tilde{\mu}_m^2\sum_{n'i'}\left[T^{(m)}_{ni,n'i'}+ V^{(m,m)}_{ni,n'i'}\right]c^{(m)}_{n'i'}
   + \sum_{n'i'} V^{(m,m+2)}_{ni,n'i'} c^{(m+2)}_{n'i'}
   + \sum_{n'i'} V^{(m,m-2)}_{ni,n'i'} c^{(m-2)}_{n'i'}
   =\tilde{M}^2\sum_{n'i'}B^{(m)}_{ni,n'i'}c_{n'i'}^{(m)},
\ee
with $\tilde{\mu}_m\equiv\mu_m\sqrt{4\pi/\lambda}$, $\tilde{M}\equiv M\sqrt{4\pi/\lambda}$,
and matrices defined as given in the appendix of \cite{phi4sympolys}.

The matrices $B^{(m)}$ come from the overlap between basis functions in each
Fock sector.  If the basis was orthonormal, $B^{(m)}$ would be the identity
matrix; however, due to round-off errors that would be associated with the construction
and use of orthonormal combinations, the basis functions are not chosen to
be orthonormal.\footnote{For low orders, orthonormal combinations become
practical because they can be constructed and used analytically, avoiding
the round-off errors associated with a numerical process.}  Instead, we
implicitly orthonormalize the basis by using a singular-value decomposition
$B^{(m)}=U^{(m)}D^{(m)}U^{(m)T}$.  The columns of the matrix $U^{(m)}$
are the eigenvectors of $B^{(m)}$.  The matrix $D^{(m)}$ is a diagonal
matrix of the eigenvalues of $B^{(m)}$.  We then define new vectors
of coefficients $\vec c^{\,(m)\prime}=D^{1/2}U^T\vec c^{\,(m)}$
and new matrices, such as $T^{(m)\prime}=D^{-1/2}U^T T^{(m)} UD^{-1/2}$,
with the $V$ matrices defined analogously.  The equations now become
\be \label{eq:reducedequations}
\tilde{\mu}_m^2\sum_{n'i'}\left[T^{(m)\prime}_{ni,n'i'}+ V^{(m,m)\prime}_{ni,n'i'}\right]c^{(m)\prime}_{n'i'}
   + \sum_{n'i'} V^{(m,m+2)\prime}_{ni,n'i'} c^{(m+2)\prime}_{n'i'}
   + \sum_{n'i'} V^{(m,m-2)\prime}_{ni,n'i'} c^{(m-2)\prime}_{n'i'}
   =\tilde{M}^2 c_{ni}^{(m)\prime},
\ee

In exact arithmetic, this transformation is well defined.  The overlap
matrix $B^{(m)}$ is a symmetric positive-definite matrix, and the 
eigenvalues must be positive, making $D^{1/2}$ real.
In practice, though, round-off error can produce
small negative eigenvalues.  Also, at high orders, some of the original
polynomials are nearly linearly dependent, which is signaled by
small positive eigenvalues.  In some sense, the basis is too large
and not fully independent.  A robust linear independence is restored
by reducing the basis size, keeping in $U^{(m)}$ only those
columns associated with eigenvalues above some positive threshold~\cite{Wilson}.
The transformation is then implicitly a projection onto a smaller basis.
For the results presented here, the threshold was $10^{-15}$, because
the need was driven by round-off errors in double-precision arithmetic.

In order to solve for $\tilde\mu_1$, we define a set of matrices $G^{(m)}$ recursively,
from $m=\Nmax$ down to 3, as
\be
G^{(m)}=\left[\tilde\mu_m^2 T^{(m)\prime}+V^{(m,m)\prime}-\tilde M^2 I^{(m)}
                -V^{(m,m+2)\prime}G^{(m+2)}V^{(m+2,m)\prime}\right]^{-1},
\ee
with the initial form given by
\be
G^{(\Nmax)}=\left[\tilde M^2 T^{(\Nmax)\prime}
                   +V^{(\Nmax,\Nmax)\prime}-\tilde M^2 I^{(\Nmax)}
                      \right]^{-1}
\ee
and $I^{(m)}$ the identity matrix in the $m$th sector.
The bare mass in the lowest sector is then simply
\be
\tilde\mu_1^2=\frac{1}{T^{(1)}}\left[\tilde M^2 -V^{(1,1)}-V^{(1,3)\prime}G^{(3)}V^{(3,1)\prime}\right],
\ee
where $T^{(1)}$ is a $1\times1$ matrix and therefore just a number.
The coefficients for the wave-function expansions are constructed recursively 
from $m=3$ up to $\Nmax$ by
\be
\vec{c}^{\,(m)\prime}/c^{(1)}=G^{(m)}V^{(m,m-2)\prime}\vec c^{\,(m-2)\prime}/c^{(1)},
\ee
with the value of $c^{(1)}$ set last by the normalization (\ref{eq:normalization}), 
which becomes
\be
1=\sum_{m=1} \vec c^{\,(m)\dagger} B^{(m)}\vec c^{\,(m)}
   =\left|c^{(1)}\right|^2+\sum_{m=3} \left|\vec c^{\,(m)\prime}\right|^2.
\ee
The values of intermediate $\mu_m$ are set by values of $\tilde\mu_1$
obtained in calculations with smaller $\Nmax$, again recursively.

As $\Nmax$ is increased, the $\tilde\mu_m$ converge to the dimensionless
bare mass $\tilde{\mu}\equiv\mu\sqrt{4\pi/\lambda}$ obtained in the standard 
parameterization, where the bare mass is not sector dependent.  This is just
the reciprocal of the dimensionless coupling $g\equiv\lambda/(4\pi\mu^2)$
used in \cite{phi4sympolys}.  This allows us to estimate $g$ as $1/\tilde\mu_1^2$.
The ratio $M^2/\mu^2$ is then obtained as $g\tilde M^2=\tilde M^2/\tilde\mu_1^2$.

\begin{figure}
\vspace{0.2in}
\centerline{\includegraphics[width=15cm]{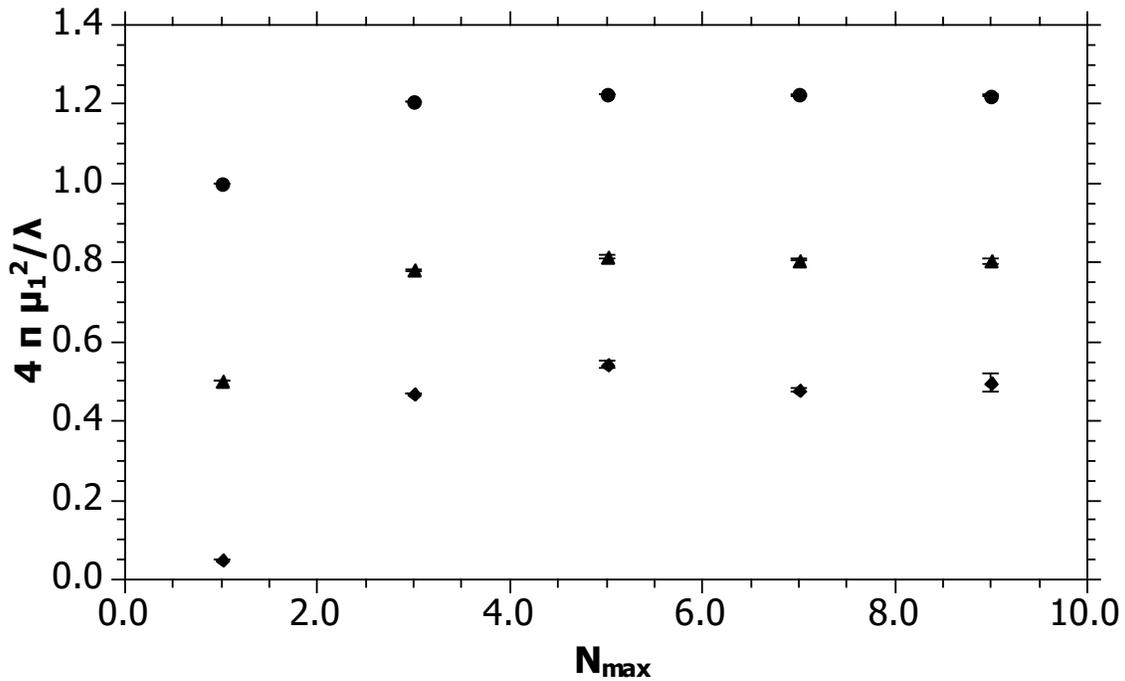}}
\caption{\label{fig:mu2vsNmax}
Plot of the scaled bare mass $\tilde\mu_1^2=4\pi\mu_1^2/\lambda$ versus
the Fock-space truncation at $\Nmax$ constituents, for three
values of $\tilde M^2$: 1.0 (circles), 0.5 (triangles), and 0.05 (diamonds).  
The error bars represent uncertainties in extrapolations.
}
\end{figure}

The convergence of $\mu_1$ is illustrated in Fig.~\ref{fig:mu2vsNmax} for
representative values of the mass scale $\tilde M^2$.  At $\Nmax=1$, the
points correspond to $\tilde\mu_1=\tilde M$; most of the change as self-energy
contributions become active occurs immediately at $\Nmax=3$.
%


\end{document}